\documentclass{aipproc}

\layoutstyle{8x11single}

\newcommand{\tltze}{$^{208}$Tl}
\newcommand{\bitof}{$^{214}$Bi}
\newcommand{\dto}{D$_{2}$O}
\newcommand{\hto}{H$_{2}$O}
\newcommand{\nhits}{N$_{hits}$}
\newcommand{\tij}{$\theta_{ij}$}

\newcommand{\nue}{$\nu_{e}$}
\newcommand{\numu}{$\nu_{\mu}$}
\newcommand{\nutau}{$\nu_{\tau}$}
\newcommand{\nux}{$\nu_{x}$}
\newcommand{\phicc}{$\phi^{\mbox{\tiny CC}}$}
\newcommand{\phies}{$\phi^{\mbox{\tiny ES}}$}

\newcommand{\nsix}{$^{16}$N}
\newcommand{\teff}{$T_{eff}$}
\newcommand{\costs}{$\cos\theta_{\odot}$}
\newcommand{\phiccsno}{$\phi^{\mbox{\tiny CC}}_{\mbox{\tiny SNO}}$}
\newcommand{\phiessno}{$\phi^{\mbox{\tiny ES}}_{\mbox{\tiny SNO}}$}
\newcommand{\phiessk}{$\phi^{\mbox{\tiny ES}}_{\mbox{\tiny SK}}$}
\newcommand{\phinumutau}{$\phi(\nu_{\mu\tau})$}

\input{epsf.tex}

\begin{document}


\title{Neutrino Observations from the Sudbury Neutrino Observatory}

\author{A.W.P. Poon 
\footnote{for the Sudbury Neutrino Observatory Collaboration: 
Q.R.~Ahmad, R.C.~Allen, T.C.~Andersen, J.D.~Anglin, 
G.~B\"{u}hler, J.C.~Barton, E.W.~Beier, 
M.~Bercovitch, J.~Bigu
% ...
}}
{
address={Institute for Nuclear and Particle Astrophysics, 
Lawrence Berkeley National Laboratory, Berkeley, CA, USA},
email={awpoon@lbl.gov}
}


\begin{abstract}
    
The Sudbury Neutrino Observatory (SNO) is a water imaging Cherenkov
detector.  Its usage of 1000 metric tons of \dto\ as target allows the
SNO detector to make a solar-model independent test of the neutrino
oscillation hypothesis by simultaneously measuring the solar \nue\
flux and the total flux of all active neutrino species.  Solar
neutrinos from the decay of $^{8}$B have been detected at SNO by the
charged-current (CC) interaction on the deuteron and by the elastic
scattering (ES) of electrons.  While the CC reaction is sensitive
exclusively to \nue, the ES reaction also has a small sensitivity to
\numu\ and \nutau.  In this paper, recent solar neutrino results from the SNO
experiment are presented.  It is demonstrated that the solar 
flux from $^{8}$B decay as measured from the ES reaction rate under 
the no-oscillation assumption is consistent with the high precision ES measurement by the 
Super-Kamiokande experiment.  The \nue\ flux deduced from the CC reaction rate in SNO
differs from the Super-Kamiokande ES results by 3.3$\sigma$.  This is evidence for
an active neutrino component, in additional to \nue, in the solar
neutrino flux.  These results also allow the first experimental
determination of the total active $^{8}$B neutrino flux from the Sun,
and is found to be in good agreement with solar model predictions.


\end{abstract}

\date{}

\maketitle

\section{Introduction}

For more than 30 years, solar neutrino
experiments~\cite{bib:homestake,bib:kamioka,bib:sage,bib:gallex,bib:gno,bib:superk}
have been observing fewer neutrinos than what are predicted by the
detailed models~\cite{bib:bpb,bib:brun} of the Sun.  A comparison of
the predicted and observed solar neutrino fluxes for these experiments
are shown in Table~\ref{tbl:solarnuexp}.  These experiments probe
different parts of the solar neutrino energy spectrum, and show an
energy dependence in the observed solar neutrino flux.  These observations can be
explained if the solar models are incomplete or neutrinos undergo
flavor transformation while in transit to the Earth.

The Sudbury Neutrino Observatory was constructed to resolve this 
solar neutrino puzzle.  It is capable of making simultaneous 
measurements of the electron-type neutrino (\nue) flux from $^{8}$B decay 
in the Sun and the flux of all active 
neutrino flavors through the following reactions:
\[\begin{array}{lcll}
    \nu_{e}+d & \rightarrow & p+p+e^{-} & \hspace{0.5in} \mbox{(CC)}\\ 
    \nu_{x}+d & \rightarrow & p+n+\nu_{x} & \hspace{0.5in} \mbox{(NC)} \\
    \nu_{x}+e^{-} & \rightarrow &  \nu_{x}+e^{-} & \hspace{0.5in} \mbox{(ES)} \\
\end{array}\]
The charged-current (CC) reaction on the deuteron is sensitive exclusively to \nue, 
and the neutral-current (NC) reaction has equal sensitivity to all 
active neutrino flavors (\nux, $x=e,\mu,\tau$).  Elastic scattering 
(ES) on electron is also sensitive to all active flavors, but with 
reduced sensitivity to \numu\ and \nutau.  

\begin{figure}
    \centering
    \epsfysize=2.5in 
    \epsfbox{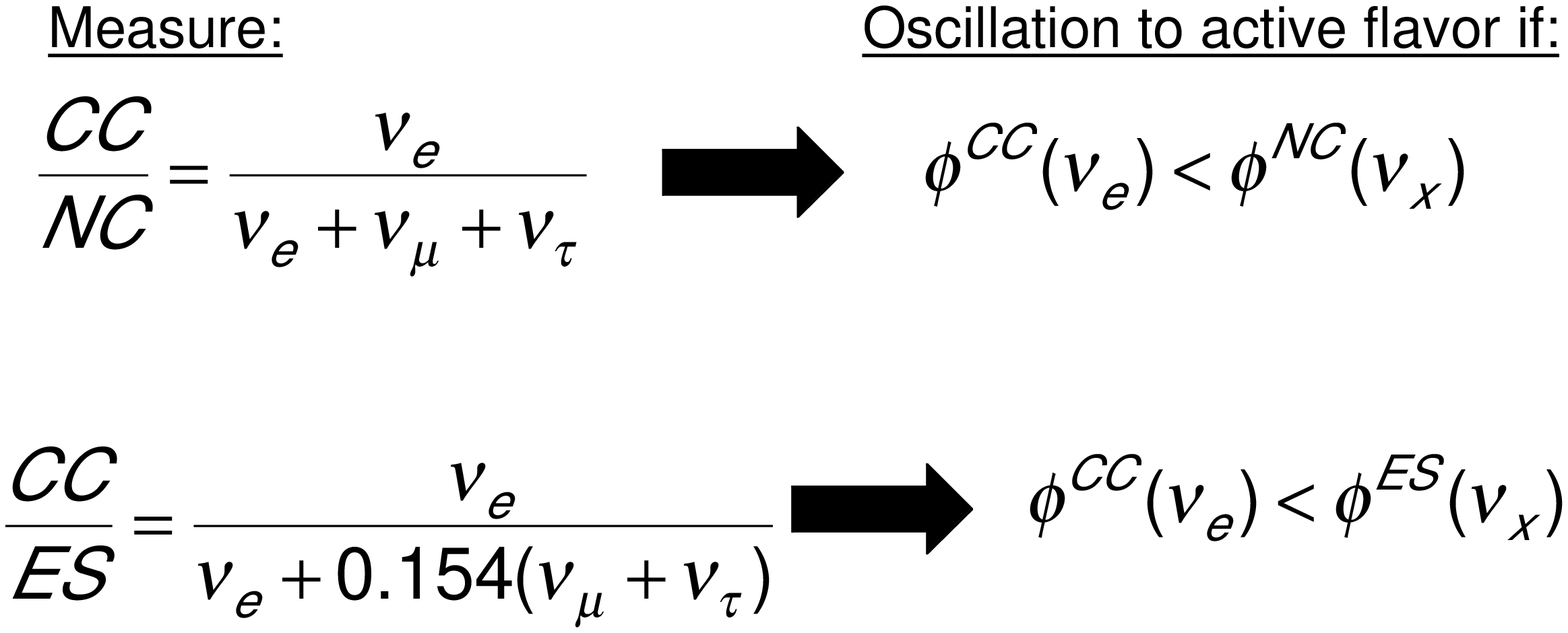}    
    \caption{Using the measured solar neutrino fluxes from different 
    reaction channels to provide ``smoking gun'' evidence of neutrino 
    oscillation.}
    \protect\label{fig:smoking_gun}
\end{figure}

Comparison of the solar neutrino flux inferred from the reaction rates
of these three reaction channels under the no-oscillation assumption
can provide ``smoking gun'' evidence for flavor-changing neutrino
oscillations.  If \nue's from the Sun transform into another active
flavor, then the solar neutrino flux deduced from the CC reaction rate
(\phicc(\nue)) must be less than those deduced from the ES reaction
rate or the NC reaction rate.  This is summarized in
Figure~\ref{fig:smoking_gun}.

Recent results~\cite{bib:snoprl} from the first measurements of the solar $^{8}$B neutrino flux by the SNO 
detector using the CC and ES reactions are presented in this 
paper.  The measured \phies(\nux) is consistent with the high precision ES 
measurement by the Super-Kamiokande Collaboration~\cite{bib:superk}.  
The measured \phicc(\nue) at SNO, however, is significantly smaller  
and is therefore inconsistent with the null hypothesis of a pure \nue\ constituent in the 
solar neutrino flux.  

\begin{table}[tp]
    \centering
    \begin{tabular}{lllll} \hline
	Experiment & Measured Flux & SSM Flux~\cite{bib:bpb} & Ref. \\ \hline
	Homestake  &  
	2.56$\pm$0.16(stat.)$\pm$0.16(sys.) SNU & 7.6$^{+1.3}_{-1.1}$SNU & 
	\cite{bib:homestake} \\ \hline
	SAGE  & 
	67.2$^{+7.2}_{-7.0}$(stat.)$^{+3.5}_{-3.0}$ SNU & 128$^{+9}_{-7}$ SNU & \cite{bib:sage} 
	\\ \hline
	Gallex & 
	77.5$\pm$6.2(stat.)$^{+4.3}_{-4.7}$(sys.) SNU & 128$^{+9}_{-7}$ SNU & \cite{bib:gallex} 
	\\ 
	GNO & 
	65.8$^{+10.2}_{-9.6}$(stat.)$^{+3.4}_{-3.6}$(sys.) SNU & 128$^{+9}_{-7}$ SNU & \cite{bib:gno} 	
	\\ \hline
	Kamiokande & 	
	2.80$\pm$0.19(stat.)$\pm$0.33(sys.)$\times$10$^{6}$~cm$^{-2}$~s$^{-1}$ & 
	5.05$\times 10^{6}\left(1^{+0.20}_{-0.16}\right)\times 10^{6}$~cm$^{-2}$~s$^{-1}$ & \cite{bib:kamioka} \\ 
	Super-Kamiokande & 	
	2.32$\pm$0.03(stat.)$^{+0.08}_{-0.07}$(sys.)$\times$10$^{6}$~cm$^{-2}$~s$^{-1}$ & 
	5.05$\times 10^{6}\left(1^{+0.20}_{-0.16}\right)\times 
	10^{6}$~cm$^{-2}$~s$^{-1}$ & \cite{bib:superk} \\ \hline

    \end{tabular}
    \caption{Summary of solar neutrino observations at different 
    solar neutrino detectors.}
    \protect\label{tbl:solarnuexp} 
\end{table}

\section{The Sudbury Neutrino Observatory}

\subsection{Physical Description of the SNO Detector}

SNO~\cite{bib:sno} is an imaging water Cherenkov detector located in
the Creighton \#9 mine, owned by the International Nickel Company
(INCO) near Sudbury, ON, Canada.  A barrel-shaped cavity with a height
of 34~m and a diameter of 22~m was excavated at a depth of 2092~m (or
6000 meters of water equivalent) to house the detector.  The muon flux
traversing the detector is 67~day$^{-1}$.

\begin{figure}
    \centering
    \epsfysize=4.5in 
    \epsfbox{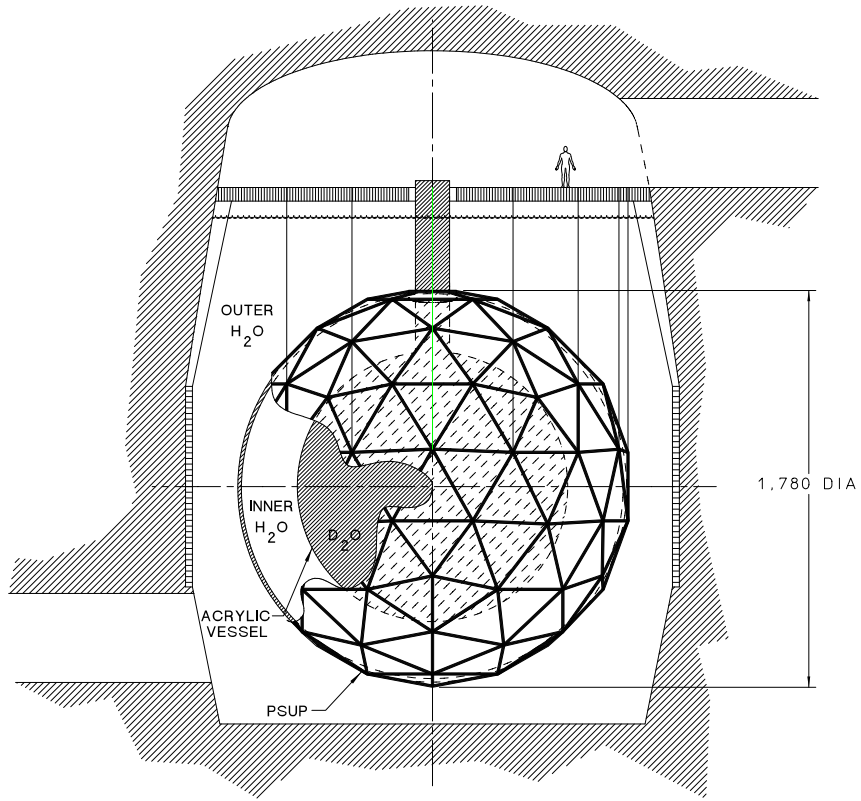}    
    \caption{A cross-sectional view of the SNO detector.  The outer 
    geodesic structure is the PMT support (``PSUP'').}
    \protect\label{fig:SNO-SK0126D_cm}
\end{figure}

Figure~\ref{fig:SNO-SK0126D_cm} shows a cross-sectional view of the
SNO detector.  The neutrino detection medium is 1000 metric tons of
99.92\% isotopically pure \dto\ contained in a 12-m diameter acrylic
sphere.  The acrylic vessel is constructed out of 122 ultraviolet
transmitting acrylic panels.  This sphere is surrounded by 7000 metric
tons of ultra-pure \hto\ contained in the cavity.  This volume of
\hto\ shields the detector from high energy $\gamma$ rays and neutrons
originating from the cavity wall.  A 17.8-m diameter stainless steel
structure supports 9456 20-cm inward-facing photomultiplier tubes
(PMTs).  A non-imaging light concentrator is mounted on each PMT, and
the total photocathode coverage is 55\%.  An additional 91 PMTs are
mounted facing outward on the support structure to serve as cosmic
veto.  To cancel the vertical components of the terrestrial magnetic
field, 14 horizontal magnetic compensation coils were built into the
cavity wall .  The maximum residual field at the PMT array is
$<$19$\mu$T, and the reduction in photo-detection efficiency is about
2.5\% from the zero-field value.

A physics event trigger is generated in the detector when there are 18 or more
PMTs exceeding a threshold of $\sim$0.25 photo-electrons within a
coincidence time window of 93~ns.  All the PMT hits registered in the
$\sim$420~ns window after the start of the coincidence time window are 
recorded in the data stream.  This widened time window allows 
scattered and reflected 
Cherenkov photons to be included in the event.  The mean noise rate 
of the PMTs is $\sim$500~Hz, which results in $\sim$2 noise PMT hits in 
this 420~ns window.  The instantaneous trigger rate is about 
15-20~Hz, of which 6-8~Hz are physics triggers.  The remaining 
triggers are diagnostic triggers for monitoring the well 
being of the detector.  The trigger efficiency reaches 100\% when the PMT
multiplicity (\nhits) in the event window is $\geq$23.

\subsection{Solar Neutrino Physics Program at SNO}

The solar neutrino physics program at SNO is designed to exploit its 
unique NC capability.  Because the result of this NC 
measurement is a definitive statement on the oscillation of solar 
neutrinos, the SNO experiment plans to make three NC measurements of the 
total $^{8}$B active neutrino flux.

The first NC measurement is made with a pure \dto\ target.  The 
free neutron from the NC interaction is thermalized, and in 
30\% of the time, a 6.25-MeV $\gamma$ ray is emitted following 
the neutron capture by deuteron.  A significant portion of the 
6.25-MeV photopeak is below the neutrino analysis threshold.  The 
second NC measurement is made with NaCl added to the \dto.  The free 
neutron is readily captured by $^{35}$Cl in this detector configuration, and a cascade 
of $\gamma$ rays with a total energy of 8.6~MeV follow.  The neutron 
detection efficiency is significantly enhanced, and $\sim$45\% of the 
NC events have a detectable signal above the analysis threshold.  
In the third NC measurement, discrete $^{3}$He proportional counters 
will be installed inside the \dto\ volume~\cite{bib:ncd}.  The neutron detection efficiency 
of the proportional counter array is 37\%.  In this detector configuration, 
the detection of the CC and the NC signals are decoupled, and the 
covariance of the CC and NC signals that appear in the first two 
detector configurations is eliminated in this case.

\section{Solar Neutrino Analysis at SNO}

\begin{figure}
    \centering
    \epsfysize=3.7in 
    \epsfbox{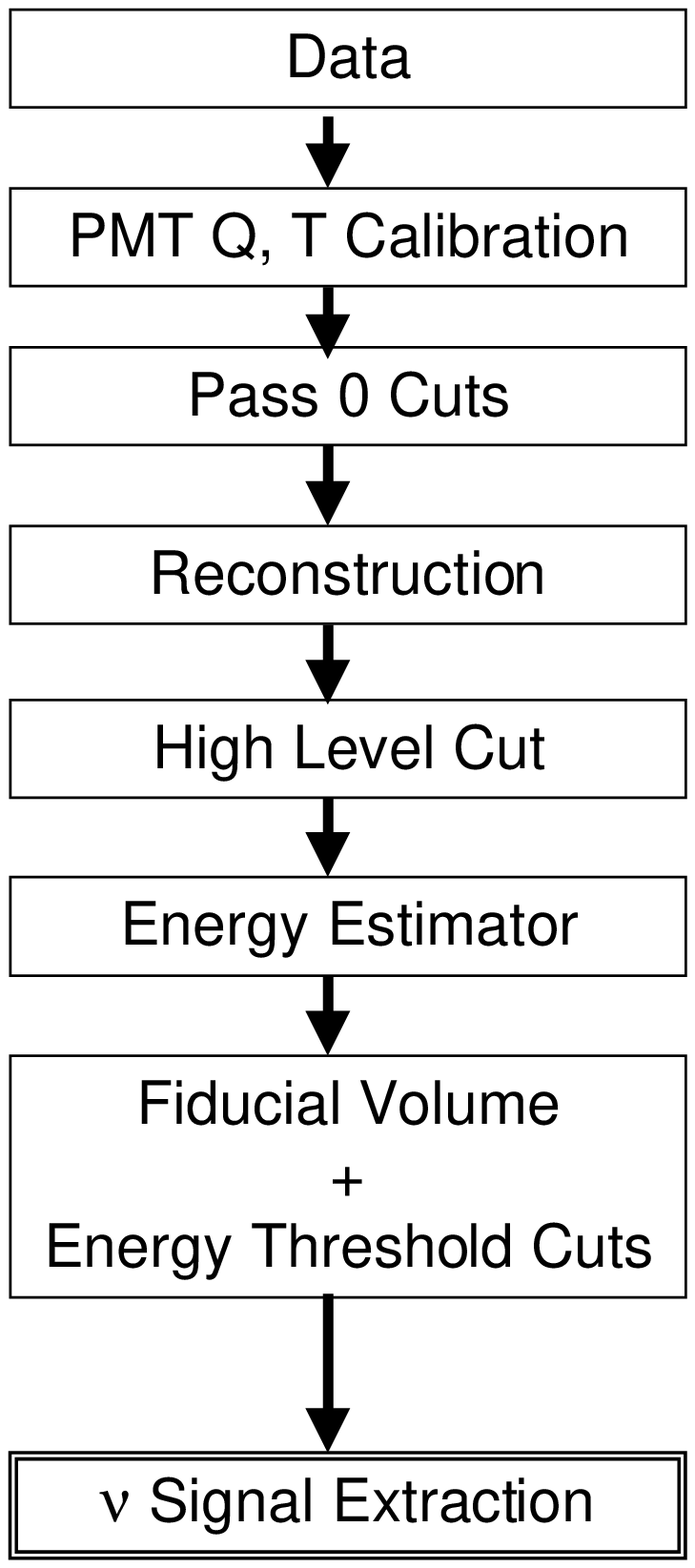}    
    \caption{Simplified flow chart of solar neutrino analysis at SNO.}
    \protect\label{fig:analysis_flow}
\end{figure}

The data presented in this paper were recorded between November 2,
1999 and January 15, 2001.   The corresponding live time is 
240.95~days.  The target was pure \dto\ during this
period.  Figure~\ref{fig:analysis_flow} summarizes the analysis procedure.  The
data were divided into two sub-sets.  One of these sub-sets contained
$\sim$70\% of the data and was used to establish the data analysis
procedures.  The remaining 30\% was used for a blind test of
statistical bias in the analysis after the analysis procedures were
settled.  Analyses of the open and the blind data sets employing the
same procedures show no statistically significant differences.  In the
following the analysis of the combined data set are presented.

\subsection{Pass 0 Cuts}

After removing all the detector diagnostic triggers from the data
stream, a significant portion of the remaining events are instrumental
backgrounds.  Electrical discharges in the PMTs (``flashers'') or
insulating detector materials emit light.  These events have
characteristic PMT time and charge distributions that are
significantly different from Cherenkov light, and can be eliminated
using cuts based on these distributions.  For example, the discharge
light emitted from a flasher PMT is detected across 
the detector $\sim$70~ns after the initial discharge is registered. 
Some of these light-emitting instrumental backgrounds are localized
near the water piping near the top of the detector.  Veto PMTs were
installed in this region in order to enhance the rejection efficiency
of these non-Cherenkov events.  Interference in the electronics system
can produce false events.  Most of the registered electronic channel
charges in these interference events are near the pedestal, and can be
removed by a cut on the mean charge of the fired PMTs.  Some of these 
electrical discharge or electronic interference background events also 
have different event-to-event time correlations from physics events, 
and time correlation cuts are used to remove these events.  Two 
independent instrumental background rejection schemes are used.  An 
event-by-event comparison of the data sets reduced by these two 
schemes shows a difference of $<$0.2\%.  

The physics loss due to these instrumental background cuts is
calibrated with a triggered \nsix\ 6.13-MeV $\gamma$-ray 
source~\cite{bib:nsix} and a
triggered $^{8}$Li 13-MeV endpoint $\beta$ source~\cite{bib:tagg} deployed to the
\dto\ and \hto\ volumes.  Further tests of the \nhits\ dependence in 
the cuts are performed with an isotropic light source at various 
intensities.  The physics acceptance of the instrumental
background cuts, weighted over the fiducial volume, is measured to be 0.9967$^{+0.0018}_{-0.0008}$.  
Instrumental background rejection is well over 99\% at this stage.  

Free neutrons and high-energy $\beta$-decay nuclei from spallations
induced by cosmic rays can form a significant background to the solar
neutrino signal.  Because the cosmic muon event rate is sufficiently
low (67~day$^{-1}$), a 20-second veto window after each cosmic muon event
is employed to eliminate any contamination of the neutrino
signal from the spallation products.

\subsection{Reconstruction}

After passing the instrumental background cuts, all events with
\nhits$\geq$30 ($\sim$3.5~MeV electron energy) are reconstructed.  The
calibrated times and positions of the fired PMTs are used to
reconstruct the vertex position and the direction of the particle. 
Two different reconstruction algorithms were developed.  An
event-by-event comparison shows excellent agreement between the data
sets reconstructed by these two algorithms.  The data presented in
this paper are reconstructed using a maximum likelihood technique
which uses both the time and angular characteristics of Cherenkov
light.  Vertex reconstruction accuracy and resolution for electrons
are measured using Compton electrons from the \nsix\ $\gamma$-ray
source, and their energy dependence is verified by the $^{8}$Li
$\beta$ source.  Compton scattered electrons from a 6.13-MeV $\gamma$
ray are preferentially scattered in the forward direction relative to
the incident $\gamma$-ray direction.  In order to minimize the effect
of finite vertex resolution on this angular resolution measurement,
only $^{16}$N events that are reconstructed to more than 150~cm from
the source are used in the measurement.  At the $^{16}$N energy
($\sim$5.5~MeV total electron energy), the vertex reconstruction
resolution is 16~cm and the angular resolution is 26.7$^{\circ}$.  
Reconstruction-related systematic uncertainties to the solar neutrino 
flux measurement is $\sim$4\%.

\subsection{Energy Estimator}

\begin{figure}
    \centering
    \epsfysize=3in 
    \epsfbox{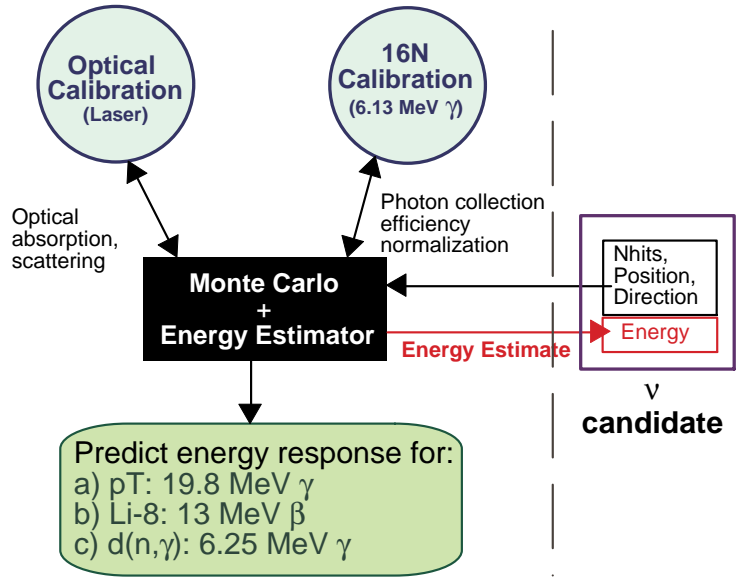}    
    \caption{Calibration of the SNO detector and event-by-event 
    energy estimator.}
    \protect\label{fig:calib_flow_chart}
\end{figure}

Figure~\ref{fig:calib_flow_chart} shows the relationship between the
detector calibration program and event-by-event energy estimation in
the analysis.  Optical calibration is obtained using a near-isotropic
source of pulsed laser light~\cite{bib:ford,bib:fordphd} at 337, 365,
386, 420, 500 and 620~nm.  The light source is deployed to locations
accessible by the source manipulator system on two orthogonal plane in
the \dto, and on a linear grid in the \hto.  Optical parameters of 
different optical media in the detector are obtained at these
wavelengths~\cite{bib:moffat}.  The attenuation lengths in \dto\ and
\hto\ are found to be near the Rayleigh scattering limit.  These
optical parameters are inputs to the Monte Carlo/energy estimator engine.

A triggered \nsix\ source (predominantly 6.13-MeV $\gamma$) is used to
provide the absolute energy calibration.  The detector energy response
to the photopeak of this source provides a normalization to the PMT
photon collection efficiency used in the Monte Carlo, and establish
the absolute energy calibration.  A long-term stability study of the
detector response to the \nsix\ source shows a linear drift of
-2.2$\pm$0.2\%~year$^{-1}$.  The cause of this effect is under
investigation, and a drift correction is applied to the event-by-event
energy estimator.

\begin{figure}
    \centering
    \epsfysize=3.5in 
    \epsfbox{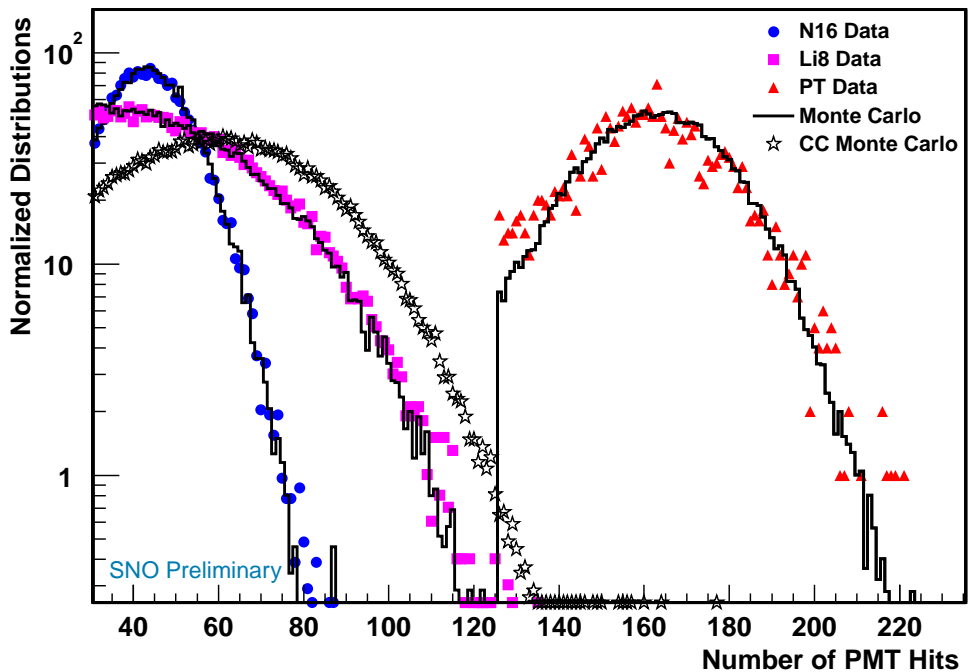}    
    \caption{Comparison of the Monte Carlo predicted responses to 
    different calibrated sources.}
    \protect\label{fig:earmycc1}
\end{figure}

This tuned Monte Carlo is then used to make predictions for the 
energy response to different calibration sources.  The pT source 
generates 19.8-MeV $\gamma$ rays through the $^{3}$H(p,$\gamma$)$^{4}$He 
reaction~\cite{bib:poon}, and is used to check the linearity of the 
energy response beyond the endpoint of the $^{8}$B neutrino energy 
spectrum.  The $^{252}$Cf fission neutron source provides an extended 
distribution of 6.25-MeV $\gamma$ rays from d(n,$\gamma$)t.  
Figure~\ref{fig:earmycc1} shows a comparison of the Monte Carlo 
predictions and the detector responses to these sources.

The energy estimator uses the same input parameters (e.g. optical
parameters) as the Monte Carlo.  It assigns an effective kinetic
energy \teff\ to each event based upon its position, direction and the
number of hit PMTs within the prompt (unscattered) photon peak.  For 
an electron of total energy $E_{e}$, the derived 
detector energy response can be parameterized by a Gaussian:
\begin{displaymath}
    R(E_{eff},E_{e})\;=\;\frac{1}{\sqrt{2\pi}\sigma_{E}(E_{e})}\exp
    \left[-\frac{1}{2}\left(\frac{E_{eff}-E_{e}}{\sigma_{E}(E_{e})}\right)^{2}\right]
\end{displaymath}
where $E_{eff}$=\teff+$m_{e}$, and the energy resolution is given by 
\begin{displaymath}
    \sigma_{E}(E_{e})=-0.4620+0.5470\sqrt{E_{e}}+0.008722 E_{e}\;\mbox{MeV}.
\end{displaymath}
The systematic uncertainty on this absolute energy calibration is 
found to be $\pm$1.4\%, which results in a neutrino flux uncertainty 
about 4 times larger. This is the most significant systematic 
uncertainty in the flux measurement.  Other energy related 
systematic uncertainties to the flux include the energy resolution 
and the energy scale linearity, and each contributes to $\leq\sim$0.5\% 
uncertainty to the flux measurement.  

A second energy estimator using \nhits\ is employed for validation
purposes.  These two energy estimators give consistent results in the
neutrino flux measurement.

\subsection{High Level Cuts}

Once the event reconstruction information becomes available after the
reconstruction, several high level physics cuts are applied to the
Pass 0-reduced data set to further reduce the instrumental 
backgrounds.  These high level cuts test the hypothesis that each 
event has the properties of electron Cherenkov light.  
The reconstruction figure-of-merit cuts test for the consistency
between the time and angular expectations for an event fitted to the 
location of the reconstructed vertex and that based on the properties of
Cherenkov light and the detector response.  

\begin{figure}
    \centering
    \epsfysize=3.5in 
    \epsfbox{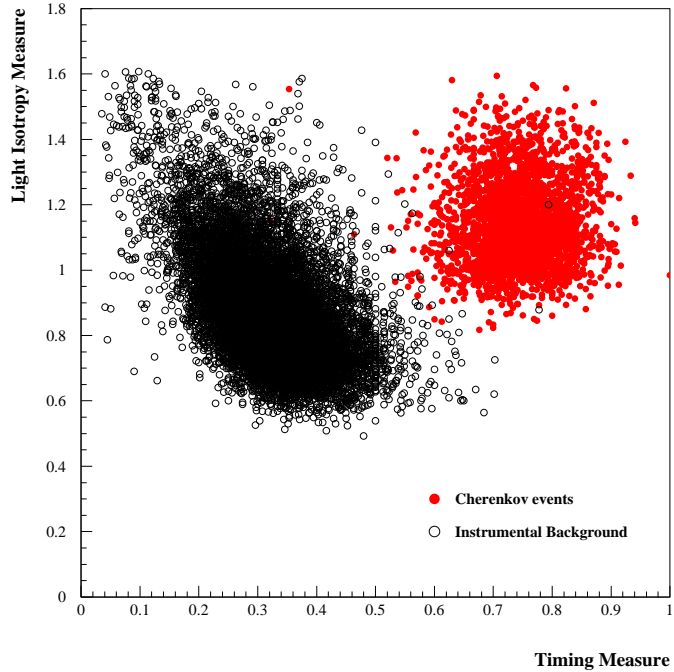}    
    \caption{Separation of instrumental backgrounds and Cherenkov 
    light events using the high level cuts.}
    \protect\label{fig:hilcuts2}
\end{figure}

Two parameters are used to further characterize Cherenkov light.  The
average opening angle between two hit PMTs ($\langle$\tij$\rangle$),
measured from the reconstructed vertex, is used to determine whether
the topology of an event is consistent with Cherenkov light.  The
in-time ratio (ITR) is the ratio of the number of hit PMTs within an
asymmetric time window around the prompt light peak to the number of
calibrated PMTs in the event.  Figure~\ref{fig:hilcuts2} shows the
correlations between \tij\ and ITR for instrumental backgrounds and
Cherenkov light events.  As shown in the figure, this two dimensional
cut has very high instrumental background rejection efficiency.

The total signal loss from the Pass 0 and the high level cuts are
calibrated with the $^{16}$N and the $^{8}$Li sources.  For the
fiducial volume (radial distance $R\leq$550~cm) and the energy
threshold (effective electron kinetic energy \teff$\geq$6.75~MeV) used
in this analysis, the volume-weighted neutrino signal loss is
determined to be 1.4$^{+0.7}_{-0.6}$\%.

The residual instrumental background contamination in the neutrino
signal after the Pass 0 and the high level cuts is estimated by a
bifurcated analysis, in which the signal contamination is obtained
from cross calibrating the background leakage of two groups of
orthogonal cuts.  For the same fiducial volume and energy thresholds,
the instrumental background contamination is estimated to be
$<$3~events (95\% C.L.), or 0.2\% of the final neutrino candidate data
set.  Table~\ref{tbl:data_reduction} summarizes the sequence of cuts 
that are used to reduce the raw data to 1169 neutrino candidate 
events. 

\begin{table}
    \centering
    \begin{tabular}{lr} \hline
	{\bf Analysis step} & {\bf Number of events} \\ \hline
	Total event triggers & 355 320 964 \\
	Neutrino data triggers & 143 756 178 \\
	PMT hit multiplicity (\nhits) $\geq$30 & 6 372 899 \\
	Instrumental background (Pass 0) cuts & 1 842 491 \\
	Muon followers & 1 809 979 \\
	High level cuts & 923 717 \\
	Fiducial volume cut & 17 884 \\
	Energy threshold cut & 1 169 \\ \hline
	{\bf Total events} & {\bf 1 169} \\ \hline
    \end{tabular}
    \caption{Data reduction steps}
    \protect\label{tbl:data_reduction}
\end{table}

\subsection{Physics Backgrounds}

\begin{figure}
    \centering
    \epsfysize=3.5in 
    \epsfbox{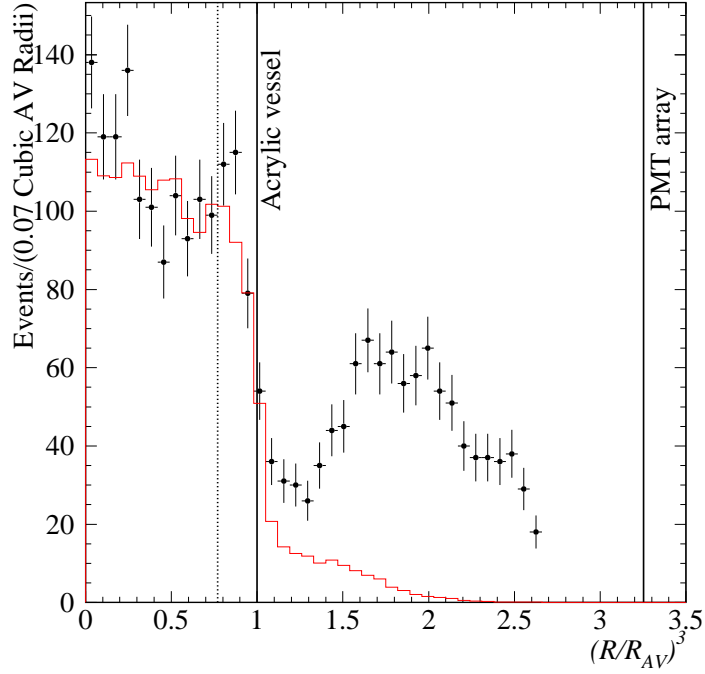}    
    \caption{Radial distribution of event candidates with 
    \teff$\geq$6.75~MeV as a function of the volume-weighted radial 
    variable $(R/R_{AV})^{3}$.  The Monte Carlo simulation of the 
    signals, weighted by the results from the signal extraction, is 
    shown as the histogram.  The dotted line indicates the fiducial 
    volume used in this analysis.}
    \protect\label{fig:fig_1}
\end{figure}

Figure~\ref{fig:fig_1} shows the radial distribution of event 
candidates with \teff$\geq$6.75~MeV as a function of the volume-weighted radial 
variable $(R/R_{AV})^{3}$, where $R_{AV}$=600~cm is the radius of the 
acrylic vessel.  Above this energy threshold, the neutrino signals 
include CC in the \dto, ES in the \dto\ and \hto, and the residual 
tail of neutron capture events (which can be NC or backgrounds from 
the photodisintegration of the deuteron), high energy tail of the 
internal radioactivity background, and high energy $\gamma$ rays 
from the cavity wall.  The simulated neutrino signals, weighted by the 
results from the signal extraction, are shown in Figure~\ref{fig:fig_1}.
The data show a clear neutrino signal within the \dto.  For the \hto\ 
region ($(R/R_{AV})^{3}>1$), the background contribution rises until 
it reaches the acceptance cutoff of the PMT light concentrators at 
$R\sim$7~m.

Detectable internal radioactivity signals are dominated by the 
$\beta\gamma$ decays of \tltze\ and \bitof, which are daughters in 
the natural Th and U chains.  These $\beta\gamma$ radionuclei can 
also emit $\gamma$ rays with sufficient energy to photodisintegrate 
the deuteron.  The free neutron from this break-up is 
indistinguishable from the NC signal.  However, this neutron background can 
be subtracted from the total neutron signal in the detector if the 
internal radioactivity level of the detector is known.  In this 
analysis, most of the Cherenkov signals from the $\beta\gamma$ decays 
are removed by the high energy threshold imposed. 
Internal radioactivity levels in the \dto\ and \hto\ are measured by regular low level 
radio-assays of U and Th chain daughters.  The light isotropy 
parameter \tij\ is also used to provide an {\em in situ} monitoring 
of the these backgrounds.  Both techniques show that the U and Th 
radioactivity levels in the \dto\ and the \hto\ are either at or 
below the target levels.  

There are also $\beta\gamma$ contributions from the construction
materials in the PMT support structure and the PMTs to the low energy
background.  Monte Carlo simulations predict that these contributions
are insignificant to the flux measurement.  This was verified by the
deployment of an encapsulated Th source in the vicinity of the PMT
support structure.  Contributions from all sources of low energy
backgrounds to the neutrino flux measurement is $<$0.2\%.

High energy $\gamma$ rays from ($\alpha,\gamma$), ($\alpha$,n$\gamma$)
and ($\alpha$,p$\gamma$) reactions in the cavity wall are
significantly attenuated by the \hto\ shield.  By deploying the \nsix\
source to the vicinity of the PMT support structure, the contribution
of these $\gamma$ rays in the event candidate set is found to be
$<$10~events (68\% CL), or a 1.9\% uncertainty to the ES flux and a 
0.8\% uncertainty to the CC flux.

\subsection{Solar Neutrino Signal Extraction}

\begin{figure}
    \centering
    \epsfysize=3.5in 
    \epsfbox{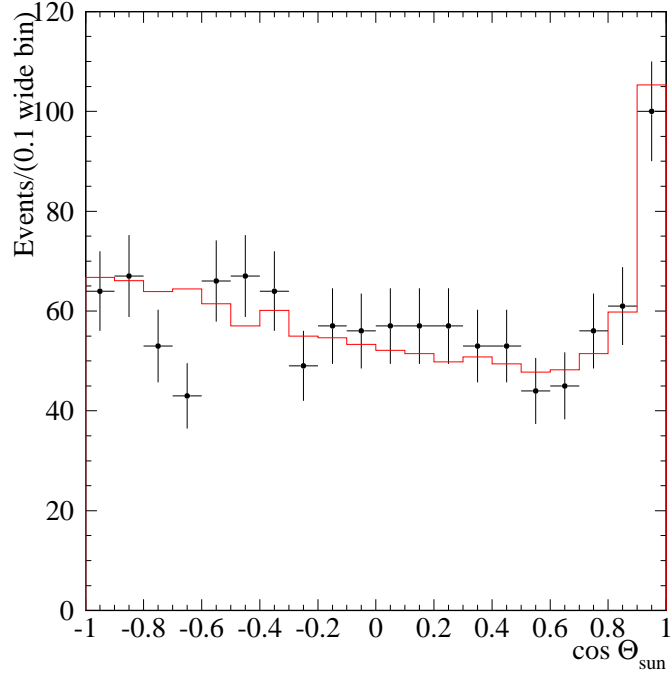} 
    \caption{\costs\ distribution
    from the candidate event set.  The Monte Carlo simulation of the
    signals, weighted by the results from the signal extraction, is
    shown as the histogram.}
    \protect\label{fig:fig2a}
\end{figure}

The extended maximum likelihood method is used in extracting the CC,
ES and neutron contributions in the candidate data set.  Data
distributions in \teff, $(R/R_{AV})^{3}$ and \costs\ are
simultaneously fitted to the probability density functions (PDFs)
generated from Monte Carlo simulations assuming no flavor
transformation and the $^{8}$B spectrum from Ortiz {\em et
al.}~\cite{bib:ortiz} \costs\ is the angle between the reconstructed 
direction of the event and the instantaneous direction from the Sun to
the Earth.  This distribution is shown in Figure~\ref{fig:fig2a}.  The
forward peak (\costs$\sim$1) arises from the strong directionality in
the ES reaction.  The \costs\ distribution for the CC reaction, before
accounting for the detector response, is expected to be
(1-0.340\costs)~\cite{bib:ebgf}.  The extraction 
yields 975.4$\pm$39.7 CC events, 106.1$\pm$15.2 ES events and 
87.5$\pm$24.7 neutron events for \teff$\geq$6.75~MeV and 
$R\leq$550~cm.  The uncertainties given above are statistical only, 
and the systematic uncertainties are summarized in 
Table~\ref{tbl:sysuncert}.

\begin{table}
    \centering
    \begin{tabular}{lrr} \hline
	Error source &  CC error &  ES error \\
	             &  (percent)&  (percent) \\ \hline
	Energy scale & -5.2,+6.1 & -3.5,+5.4 \\
	Energy resolution & $\pm$0.5 & $\pm$0.3 \\
	Energy scale non-linearity & $\pm$0.5 & $\pm$0.4 \\
	Vertex accuracy & $\pm$3.1 & $\pm$3.3 \\
	Vertex resolution & $\pm$0.7 & $\pm$0.4 \\
	Angular resolution & $\pm$0.5 & $\pm$2.2 \\
	High energy $\gamma$'s & -0.8,+0.0 & -1.9,+0.0 \\
	Low energy background & -0.2,+0.0 & -0.2,+0.0 \\
	Instrumental background & -0.2,+0.0 & -0.6,+0.0 \\
	Trigger efficiency & 0.0 & 0.0 \\
	Live time & $\pm$0.1 & $\pm$0.1 \\
	Cut acceptance & -0.6,+0.7 & -0.6,+0.7 \\
	Earth orbit eccentricity & $\pm$0.1 & $\pm$0.1 \\
	$^{17}$O, $^{18}$O & 0.0 & 0.0 \\ \hline
	{\bf Experimental uncertainty} & {\bf -6.2,+7.0} & {\bf -5.7,+6.8} \\ \hline
	Cross section & 3.0 & 0.5 \\
	Solar Model & -16,+20 & -16,+20 \\ \hline
    \end{tabular}
    \caption{Systematic uncertainties on fluxes}
    \protect\label{tbl:sysuncert}
\end{table}

The $^{8}$B neutrino flux can be determined from normalizing the 
observed integrated event rate above the energy threshold.  Assuming 
the $^{8}$B spectrum from Ref.~\cite{bib:ortiz}, the flux deduced 
from the CC and the ES reactions are:
\[\begin{array}{lcl}
    \phi^{\mbox{\tiny CC}}_{\mbox{\tiny SNO}}(\nu_{e}) &=& 
    1.75\pm 0.07\mbox{(stat.)}^{+0.12}_{-0.11}\mbox{(sys.)}\pm 
    0.05\mbox{(theor.)}\times 10^{6}\;\mbox{cm}^{-2}\mbox{s}^{-1}  \\ 
    \phi^{\mbox{\tiny ES}}_{\mbox{\tiny SNO}}(\nu_{x}) &=& 
    2.39\pm 0.34\mbox{(stat.)}^{+0.16}_{-0.14}\mbox{(sys.)}\times 
    10^{6}\;\mbox{cm}^{-2}\mbox{s}^{-1} \\ 
\end{array}\]
where the theoretical uncertainty is the CC cross section 
uncertainty~\cite{bib:ccuncert}.  Radiative corrections to the CC 
cross section have not been applied to the CC cross section, but they 
are expected to decrease the measured \phiccsno(\nue) by up to a few 
percent~\cite{bib:radcorr}.  The difference between \phiccsno\ and 
\phiessno\ is 0.64$\pm$0.40$\times$10$^{6}$~cm$^{-2}$s$^{-1}$, or 
1.6$\sigma$.  The ratio of \phiccsno\ to the predicted $^{8}$B solar 
neutrino flux from BPB01 solar model~\cite{bib:bpb} is 0.347$\pm$0.029 
where all the uncertainties are added in quadrature.  Independent 
analyses using \nhits\ as energy estimator, or in various fiducial 
volumes up to 620~cm with the inclusion of background PDFs in the signal extraction 
give consistent results.

The Super-Kamiokande experiment has made a high precision measurement 
of the $^{8}$B solar neutrino flux by the ES reaction:
\begin{displaymath}
        \phi^{\mbox{\tiny ES}}_{\mbox{\tiny SK}}(\nu_{x}) \;=\; 
    2.32\pm 0.03\mbox{(stat.)}^{+0.08}_{-0.07}\mbox{(sys.)}\times 
    10^{6}\;\mbox{cm}^{-2}\mbox{s}^{-1}. \\ 
\end{displaymath}
\phiessno(\nux) and \phiessk(\nux) are consistent with each 
other.  Assuming that the systematic uncertainties are normally 
distributed, the difference is 
\begin{displaymath}
   \phi^{\mbox{\tiny ES}}_{\mbox{\tiny SK}}(\nu_{x})-\phi^{\mbox{\tiny 
   CC}}_{\mbox{\tiny SNO}}(\nu_{e}) \;=\;
            0.57\pm 0.17 \times 10^{6}\;\mbox{cm}^{-2}\mbox{s}^{-1}, 
\end{displaymath}
or 3.3$\sigma$.  The probability that the observed \phiccsno(\nue) is 
a $\geq$3.3$\sigma$ downward fluctuation of \phiessk(\nux) is 0.04\%.  

If \nue's from $^{8}$B decays in the Sun oscillate exclusively 
to sterile neutrinos, the SNO CC-derived $^{8}$B flux with \teff$\geq$6.75~MeV 
would be consistent with the integrated Super-Kamiokande ES-derived 
flux above a threshold of 8.5~MeV~\cite{bib:fogli}.  The difference 
between these derived fluxes after adjusting for the ES 
threshold~\cite{bib:superk} is 0.53$\pm$0.17$\times 
10^{6}$~cm$^{-2}$s$^{-1}$, or 3.1$\sigma$.  The probability of a 
$\geq$3.1$\sigma$ downward fluctuation  is 0.13\%.  Therefore, the results 
presented here are evidence for a non-electron type active neutrino 
component in the solar neutrino flux.

\begin{figure}
    \centering
    \epsfysize=3in 
    \epsfbox{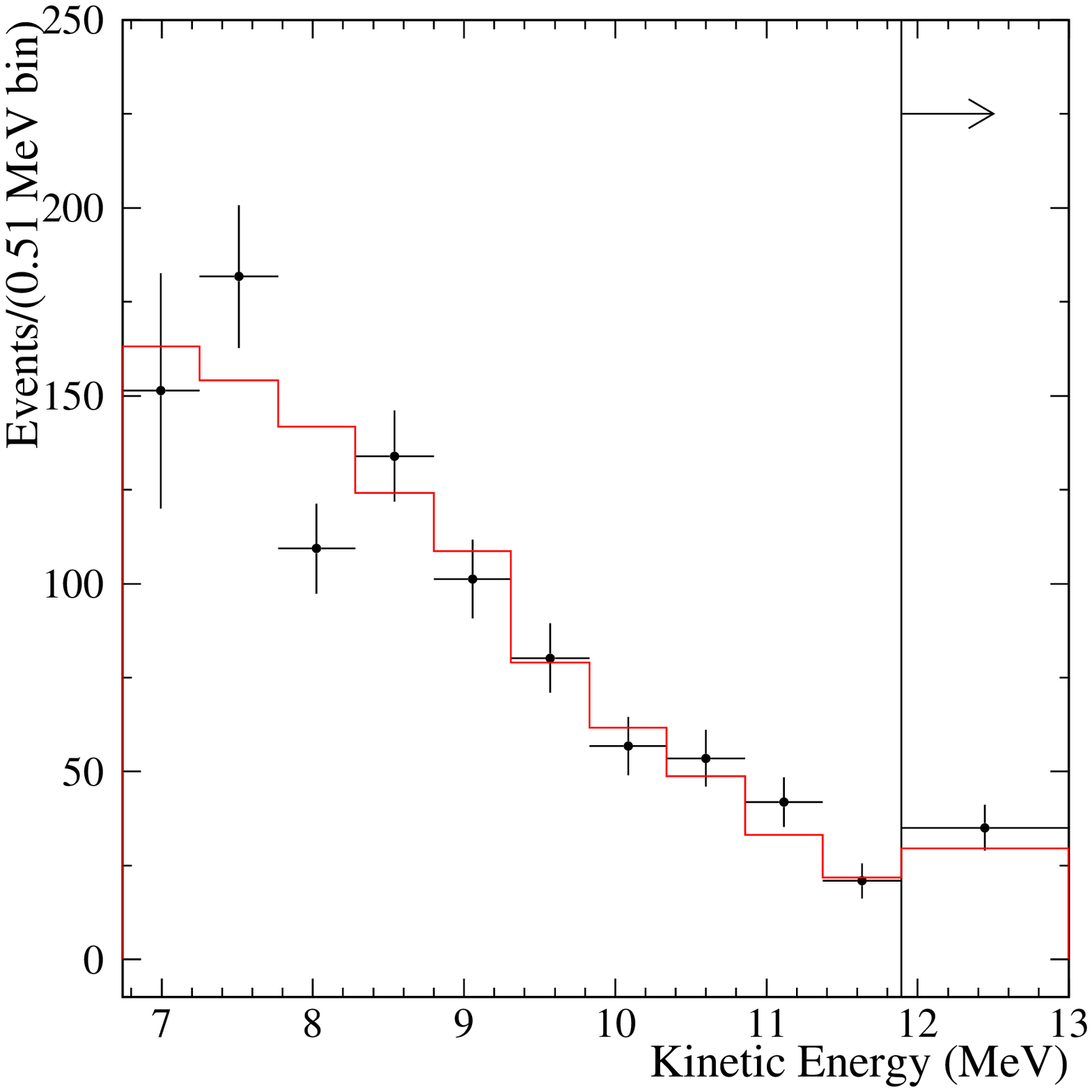} 
    \epsfysize=3in 
    \epsfbox{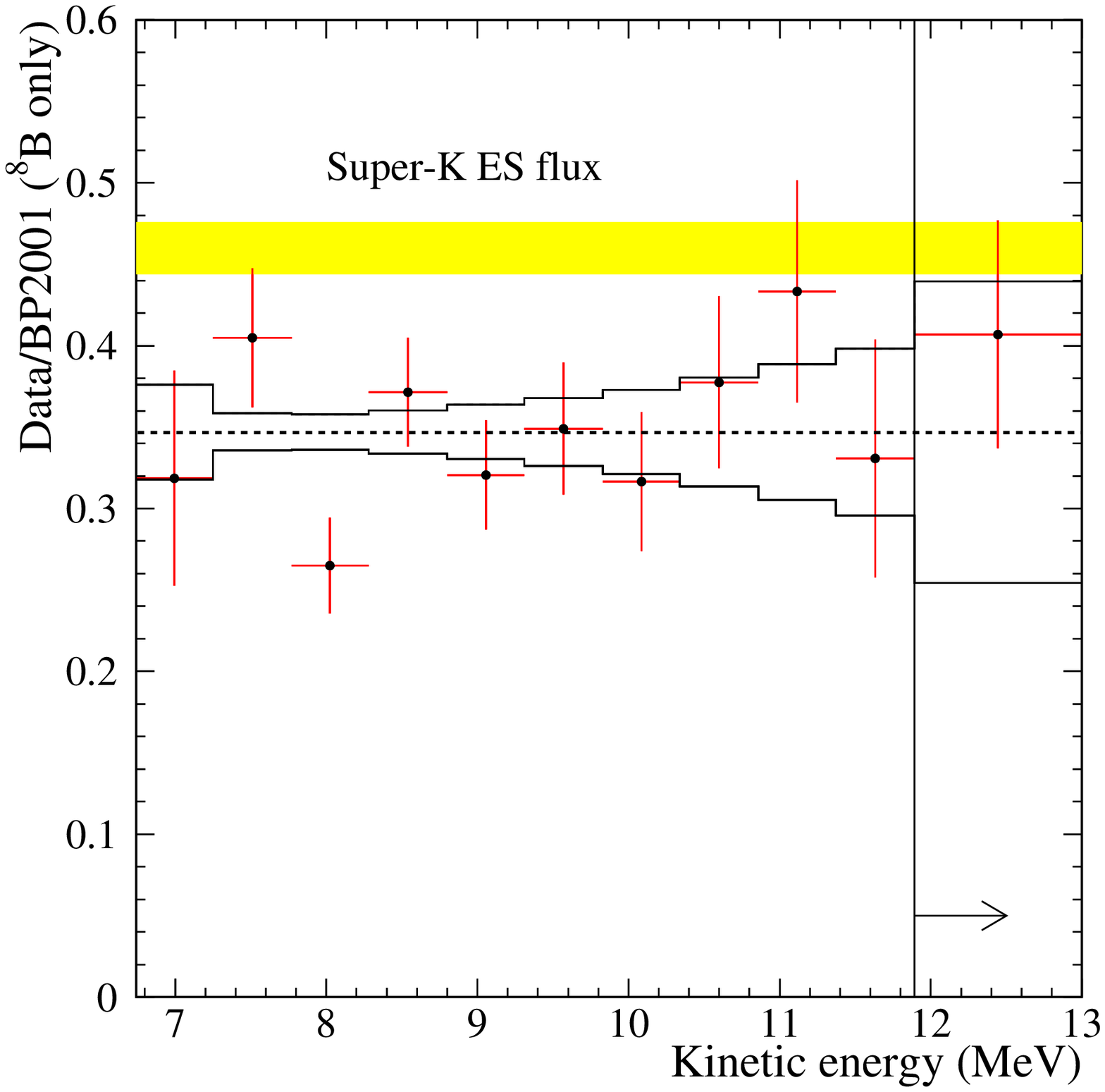}
    \caption{SNO CC energy spectrum.  {\em Left:} The extracted CC
    kinetic energy spectrum from a shape-unconstrained fit of events
    with $R\leq$550~cm and \teff$\geq$6.75~MeV. The error bars are
    statistical only.  The expected undistorted $^{8}$B spectrum,
    derived from Ref.~\cite{bib:ortiz}, is shown as the histogram. 
    {\em Right:} The ratio of the extracted CC spectrum to the
    expected kinetic energy spectrum.  The band at each energy bin
    represent the 1$\sigma$ uncertainty derived from the most
    significant energy-dependent systematic uncertainties.  The
    uncertainties in the $^{8}$B spectrum have not been included.}
    \protect\label{fig:sno_egy_spec}
\end{figure}

The CC energy spectrum can be extracted from the data by repeating 
the signal extraction with the CC energy spectral constraint 
removed.  This is shown in Figure~\ref{fig:sno_egy_spec}.  There is 
no evidence for spectral distortion under the no-oscillation hypothesis.

\section{Total Active $^{8}$B Neutrino Flux}

\begin{figure}
    \centering
    \epsfysize=3.5in 
    \epsfbox{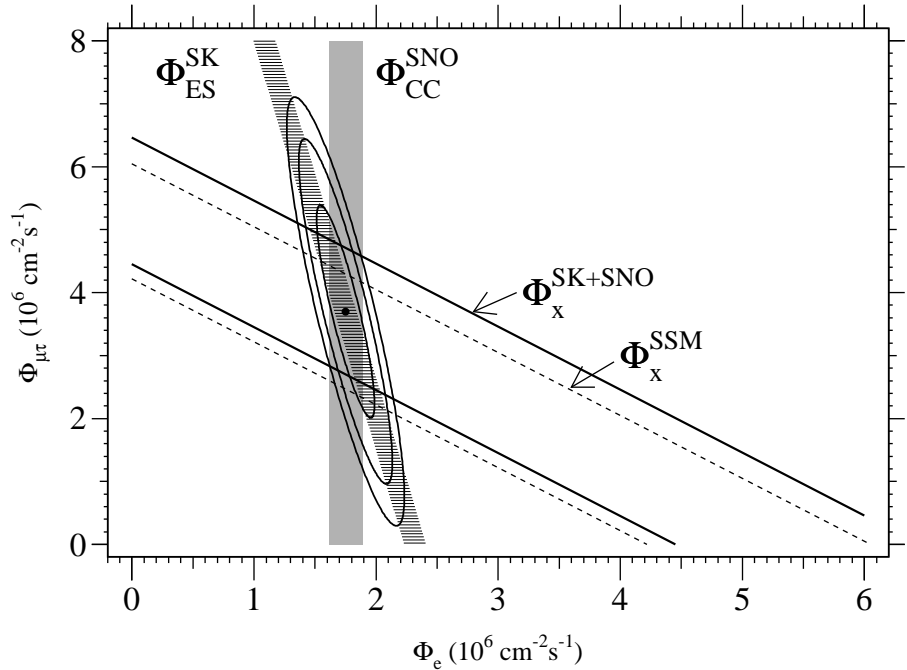} 
    \caption{Flux of non-electron flavor active $^{8}$B solar 
    neutrinos (\phinumutau) versus \phiccsno\ and \phiessk.  The band 
    derived from the SNO and the Super-Kamiokande results 
    ($\phi^{\mbox{\tiny SK+SNO}}_{x}$) and the BPB01 prediction 
    ($\phi^{\mbox{\tiny SSM}}_{x}$) are in good agreement.}
    \protect\label{fig:fig_3}
\end{figure}

Recall from Figure~\ref{fig:smoking_gun} that the ES reaction is sensitive to all active neutrino 
flavors, but with reduced sensitivity for \numu\ and \nutau.  Using the 
high precision ES measurement \phiessk(\nux) and the pure \nue\ flux 
from \phiccsno(\nue), one can infer the flux of non-electron flavor 
active neutrino \phinumutau:
\begin{displaymath}
        \phi^{\mbox{\tiny ES}}_{\mbox{\tiny SK}}=\phi(\nu_{e})+0.154\phi(\nu_{\mu\tau}).
\end{displaymath}
This is shown in Figure~\ref{fig:fig_3}, in which \phinumutau\ is 
shown against $\phi(\nu_{e})$.  The two data bands are 1$\sigma$ 
measurements of \phiccsno\ and \phiessk, and the error ellipses are 
68\%, 95\% and 99\% joint probability contours for $\phi(\nu_{e})$ 
and \phinumutau.  The best fit to \phinumutau is
\begin{displaymath}
        \phi(\nu_{\mu\tau})\;=\;3.69\pm 1.13 \times 
        10^{6}\,\mbox{cm}^{-2}\mbox{s}^{-1}. 
\end{displaymath}

The total $^{8}$B flux derived from the SNO and the Super-Kamiokande
experiments is shown as the diagonal band ($\phi^{\mbox{\tiny
SK+SNO}}_{x}$) in Figure~\ref{fig:fig_3}.  The agreement with the
standard solar model prediction ($\phi^{\mbox{\tiny SSM}}_{x}$) is
good.  The total flux of of active $^{8}$B neutrinos is found to be
\begin{displaymath}
        \phi(\nu_x)\;=\;5.44\pm 0.99 \times 
        10^{6}\,\mbox{cm}^{-2}\mbox{s}^{-1},
\end{displaymath}
This is the first determination of the total flux of $^{8}$B neutrinos generated by the 
Sun.

\section{Summary and Outlook}

Two significant results are reported in this paper.  The data from 
SNO represent the first direct evidence that there is an active
non-electron flavor neutrino component in the solar neutrino flux, and
an exclusive oscillation from \nue's to sterile neutrinos is
disfavored at the 3.1$\sigma$ level.  This is also the first
experimental determination of the total flux of active $^{8}$B solar
neutrinos, which is in good agreement with the solar model
predictions.

The SNO Collaboration is now analyzing the data from the pure \dto\ 
phase with a lowered energy threshold.  Efforts are devoted to 
understanding the low energy $\beta\gamma$ and the 
photodisintegration contribution to the NC measurement.  Results from 
this analysis will be reported in the near future.

\begin{figure}
    \centering
    \epsfysize=3in 
    \epsfbox{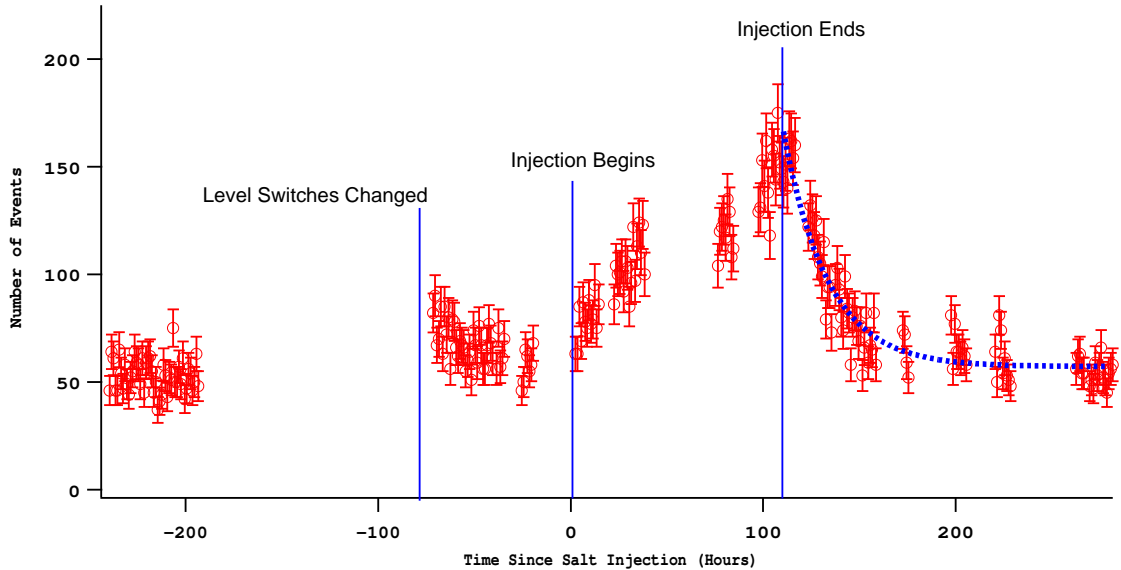} 
    \caption{Event rate in a
    low energy background monitoring window before, during and after
    the NaCl injection.  The gaps in this plot indicate detector down
    time, when detector hardware changes necessary to salt
    deployment were made.  The dotted line is an exponential fit of the 
    event rate.  The fit is consistent with $\tau_{1/2}$=15~hours, 
    which is the half life of $^{24}$Na.}
    \protect\label{fig:salt_time}
\end{figure}

The SNO experiment has just finished the first phase of the experiment.  The
deployment of NaCl to enhance the NC capability began on May 28, 2001. 
Figure~\ref{fig:salt_time} shows the detector background level seen in
the Cherenkov data before, during and after the NaCl injection.  The
increase in the event rate during the injection is attributed to
$^{24}$Na, which were activated by neutrons from the cavity wall
when the NaCl brine was stored in the underground laboratory prior to
the injection.  After the injection has ended, one sees the decay of
$^{24}$Na with a characteristic $\tau_{1/2}$=15~hours.  The
background level in the detector returned to the pre-injection level
after several days.  After 8 months of running in this configuration, 
SNO will be able to make a definitive, solar model-independent statement (better than 
6$\sigma$) on the solar neutrino oscillation hypothesis.

\section{Acknowledgements}

This research was supported by the Natural Sciences and Engineering 
Research Council of Canada, Industry Canada, National Research 
Council of Canada, Northern Ontario Heritage Fund Corporation and the 
Province of Ontario, the United States Department of Energy and in the 
United Kingdom by the Science and Engineering Research Council and the 
Particle Physics and Astronomy Research Council.  Further support was 
provided by INCO, Ltd., Atomic Energy of Canada Limited (AECL), 
Agra-Monenco, Canatom, Canadian Microelectronics Corporation, AT\&T 
Microelectronics, Northern Telecom and British Nuclear Fuels, Ltd.  
The heavy water was loaned by AECL with the cooperation of Ontario 
Power Generation.

\end{document}